# Suppression of off-resonant cavity feeding through quasi-resonant pumping in a strongly coupled cavity-quantum dot system.

**Deepak Sridharan,**[1,2,3] **Ranojoy Bose,**[1,2,3§] **Glenn S. Solomon**[2] **and Edo Waks**[1,2*]

[1]*Department of Electrical Engineering & Institute for Research in Electronics and Applied Physics, University of Maryland, College Park, MD 20742, USA*
[2]*Joint Quantum Institute, University of Maryland, College Park, MD 20742 and National Institute of Standards and Technology, Gaithersburg, MD, USA*
[3]*These authors contributed equally.*
[*]*edowaks@umd.edu*, [§]*rbose@umd.edu*

**Abstract:** We compare the photoluminescence spectrum of an indium arsenide (InAs) quantum dot (QD) that is strongly coupled to a photonic crystal cavity under above band excitation (ABE) and quasi-resonant excitation (QRE). We show that off-resonant cavity feeding, which manifests itself in a bare cavity emission peak at the strong coupling point, is suppressed by as much as 40% under QRE relative to ABE. We attribute this suppression to a reduced probability of QD charging because electrons and holes are created in pairs inside the QD. We investigate the pump power dependence of the cavity feeding and show that, below saturation, the ratio of the bare cavity emission to polariton emission for ABE is independent of pump power, while for QRE there is linear pump power dependence. These results suggest that the biexciton plays an important role in cavity feeding for QRE.

**1. Introduction**

Solid-state systems comprising of semiconductor quantum dots (QDs) coupled to photonic crystal nanocavities are promising candidates for a variety of applications in quantum information processing [1-5] and nonlinear optics at the single photon level [6-7], and can also serve as an effective spectroscopic tool for studying the fine structure of QD emission [8]. The design of high-quality factor ($Q$) photonic crystal resonators, combined with improved fabrication and lithography techniques, has resulted in the achievement of strong-coupling between photonic crystal cavities and self-assembled semiconductor QDs [6, 8-15]. In the standard theory of strong coupling between a cavity and QD, the cavity spectrum is predicted to split into two dressed polariton states, separated by twice the exciton-photon coupling strength ($g$). However, recent measurements have reported the presence of an additional peak corresponding to emission from the bare (uncoupled) cavity in the spectrum of a strongly coupled QD-cavity system on resonance [8, 10, 16]. Recent theoretical work has shown that this anomaly is well explained by other optically active states of the QD, such as the charged exciton and biexciton [17]. When the QD falls into one of these alternate charge configurations it becomes resonantly detuned from the cavity, and is therefore no longer strongly coupled. However, the detuned QD can still feed the bare cavity mode, an effect that has been experimentally observed through direct pumping of the QD state [18-19] as well as second order correlation measurements [16, 20], and theoretically explained to be caused by pure dephasing in the QD [17, 21].

Observation of bare cavity emission in the strong coupling regime is expected to be strongly dependent on the method used to optically excite the QD. Previous experiments that observed the spectral "triplet" (consisting of the dressed polariton and bare cavity states) employed above-band laser excitation (ABE) of the QD, where electrons and holes are created in the barrier material (or alternately in the wetting layer), and are randomly captured into the QD. Under these excitation conditions, there is a significant probability of creating a charged QD through capture of only a single carrier, as observed in direct spectroscopic measurements of isolated bare dots [20, 22-24]. One method to reduce the charging of the QD is to excite through a higher state [22-24], a method that we refer to as quasi-resonant excitation (QRE). Under QRE, electrons and holes are created in pairs inside the QD, significantly reducing the probability of charging. The carriers relax to the ground state through phonon mediated relaxation on the time scale of 10 ps. Reduction of charging with QRE of bare QDs has been experimentally observed [22-23].

In this paper we compare the emission of an indium arsenide (InAs) quantum dot coupled to a photonic crystal cavity in the strong coupling regime under both ABE and QRE. We observe significant suppression of off-resonant cavity feeding under QRE as compared to ABE which manifests itself in reduction of the bare cavity emission peak at strong coupling. This suppression results in an emission spectrum that is much more consistent with the standard theory of strong coupling between a cavity and a two-level emitter. We study the pump-power dependence of the emission spectrum at strong coupling using both excitation methods. We show that below saturation the relative contribution of the cavity feeding to the emission spectrum for ABE is nearly independent of pump power, which is consistent with recent theoretical predictions [17]. For QRE a linear pump-power dependence is observed below saturation, suggesting that the cavity feeding is strongly driven by biexciton emission. These results provide important insight into the underlying mechanisms for the cause of the spectral triplet in cavity coupled InAs QDs, and provide a better physical understanding of QD-cavity interactions with multiple excitonic configurations.

## 2. Fabrication and optical characterization of photonic crystal devices

Quantum dot samples used in this work were grown by molecular beam epitaxy (MBE). The initial wafer is composed of a 160-nm gallium arsenide (GaAs) membrane with an InAs QD layer grown at the center (the QD density is ~50 $\mu m^{-2}$), on a 1-μm thick sacrificial layer of $Al_{0.78}Ga_{0.22}As$. The bulk photoluminescence spectrum indicates that the QD emission is largely concentrated in the 900-960 nm wavelength range. Photonic crystals are defined on the GaAs membrane using electron-beam lithography (RAITH eLine) (Figure 1a) and chlorine-based inductively coupled plasma dry etching, followed by a selective wet etch to remove the sacrificial AlGaAs layer, resulting in a free-standing GaAs membrane. The cavity design used in this experiment is a three-hole linear defect (L3) cavity with three-hole tuning [25]. Holes A, B, and C (indicated in Fig. 1a) are shifted by 0.176*a*, 0.024*a* and 0.176*a*, where the lattice parameter *a* is 240 nm, and the hole-radii are fixed at 130 nm. Fabricated structures are placed in a continuous-flow liquid He cryostat and cooled to a temperature ranging between 8 and 40K. QDs in the cavity region are excited by a wavelength tunable Ti:Sapphire laser with a tuning range of 700-1000 nm. Emission is collected by a confocal microscope setup using a 0.7 NA objective lens. The collected emission is then measured by a spectrometer with a wavelength resolution of 0.02 nm.

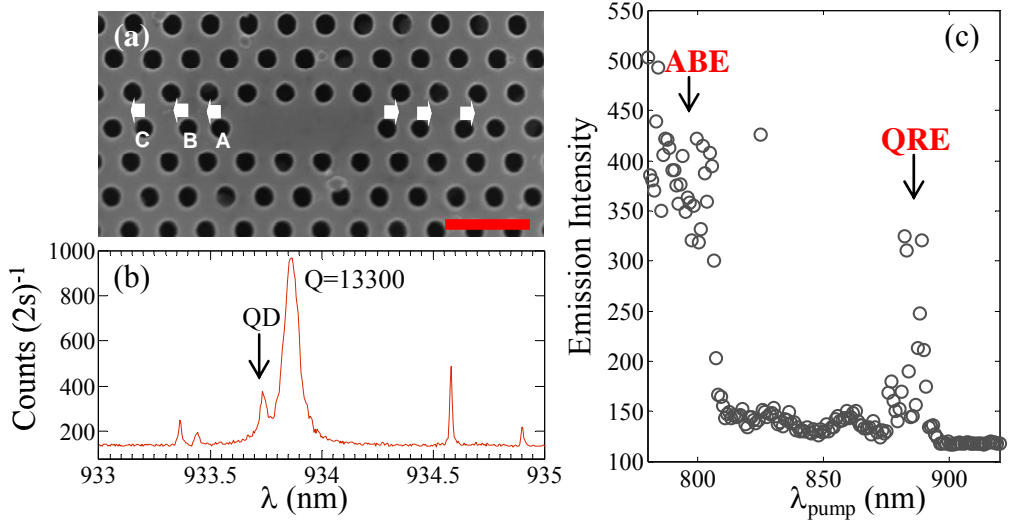

Fig. 1. (a) Scanning electron micrograph of a fabricated photonic crystal cavity with three-hole (A, B, C) tuning as indicated in text. Scale bar: 500 nm; (b) PL from the cavity mode and nearby QD states away from the strong coupling regime after cavity has been nitrogen shifted to 933.9 nm. The line labeled "QD" strongly couples to the cavity. (c) Photoluminescence excitation (PLE) spectrum showing cavity mode intensity as a function of pump wavelength $\lambda_{pump}$ (10 μW) of tunable Ti:Sapphire laser near the resonance (strong-coupling) condition, showing a clear resonance feature at 877 nm.

For this device, the QD emission wavelength was initially red-detuned from the cavity mode at 15 K. We therefore used nitrogen gas deposition [26] to red shift the cavity mode to slightly longer wavelength than the QD, and then used more detailed temperature tuning near resonance. Nitrogen deposition allows large wavelength tuning of only the cavity mode (in excess of 4 nm redshifts), while temperature-tuning allows for repeated measurements over a short wavelength range (1nm). Fig. 1b shows a photoluminescence spectrum of the fabricated device under low excitation powers, taken at 15K, after the cavity has been nitrogen shifted to $\lambda$ = 933.9 nm. From the cavity spectrum the $Q$ is calculated to be 13,300. We also indicate the strongly coupled QD that we study in this paper, which is blue shifted by 0.15 nm after gas deposition.

To identify the QRE pump wavelength we first use nitrogen gas deposition to shift the cavity on resonance with the QD. We then pump the QD with a tunable Ti:Sapphire laser over the wavelength range of 780-920 nm while observing the emission from the coupled cavity-QD spectrum in a photoluminescence excitation (PLE) measurement. Fig. 1c shows the measured PLE spectrum. High-energy excited states of a QD are observed as narrow spectral windows of enhanced QD photoluminescence (PL). In some devices we are able to observe two excited states, approximately 30 nm and 55 nm detuned from the QD emission state, while in other devices we only observe one excited state at 55 nm detuning. For the device used in these experiments, only the 55-nm detuned second excited state (877 nm) is able to efficiently drive the QD, as can be seen in Fig. 1c. The fact that considerable photoluminescence is observed under QRE indicates that the QD emission at $\lambda$ = 933.75 nm corresponds to a neutral exciton transition.

## 3. Comparison of above-band and quasi-resonant pumping

A comparison of the emission spectra under ABE and QRE for the coupled cavity-QD system is shown in Fig. 2. These measurements are obtained by first performing nitrogen gas deposition until the cavity resonance is shifted to $\lambda_c$=933.9 nm, slightly red-shifted from the neutral exciton emission at 8 K. Temperature tuning is then used to accurately shift the QD across the cavity resonance. Fig. 2a shows the photoluminescence spectrum under ABE, where the pump wavelength is set to 790 nm and the pump power is set to 1 µW (which is below QD saturation), for several different temperatures. The cavity-QD system exhibits significant cavity feeding under ABE, which overwhelms the emission from the two polariton states making them barely discernible. In Fig. 2b we plot the photoluminescence spectrum under QRE. The pump wavelength is set to 877 nm and the pump power to 4 µW (also under QD saturation). Under QRE the bare cavity emission is significantly suppressed and the emission spectrum appears much closer to the spectrum predicted by the standard theory of strong coupling between a cavity and a two-level system. From the QRE spectrum on resonance the splitting between the two polariton peaks is measured to be 0.11 nm, which corresponds to a vacuum Rabi frequency of g=18.9 GHz. Fig. 2c plots the photoluminescence spectra for ABE and QRE at the strong coupling point on top of each other, normalized to the emission peak of the lower polariton. The reduction in the bare cavity emission is clear.

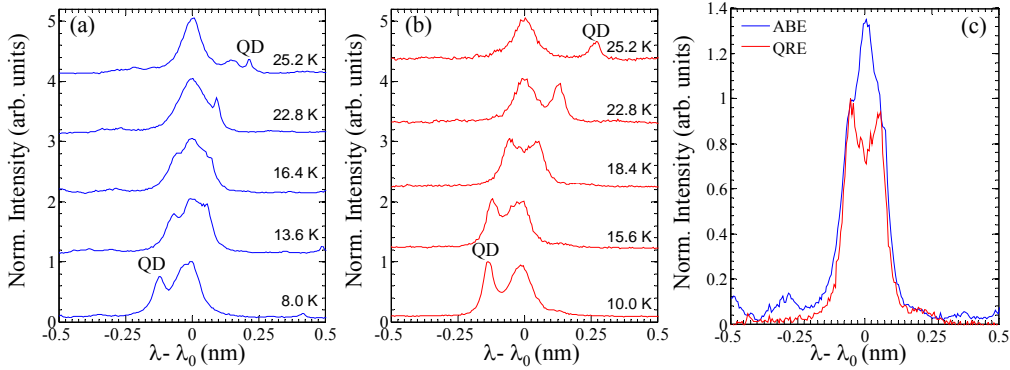

Fig. 2. ABE and QRE scanning of QD-photonic crystal nanocavity system. (a) ABE at 790 nm pump wavelength and 1µW power. Strong coupling is established at 16.4 K by tuning the QD near the cavity resonance. The cavity emission is given by $\lambda_0$=933.8 nm. (b) QRE with 877 nm pump wavelength and 4 µW pump power. Strong coupling is established at 18.4 K. (c) Comparison of ABE and QRE at the strong coupling point showing clear suppression due to QRE when normalized to the polariton emission.

The amount of bare cavity emission at the strong coupling point in general depends on the pumping intensity. In order to make quantitative statements about the degree of suppression of bare-cavity emission, it is therefore crucial to investigate the pump power dependence of the photoluminescence spectrum. We measure the cavity-QD spectra at the strong coupling point for various values of the pump power below and approaching QD saturation in both ABE and QRE. For each value of the pump power we fit the recorded PL spectrum using a three Lorentzian model (one each for the lower- and upper polaritons and the bare cavity), given by:

$$I(\lambda) = A_{lower}\, L(\gamma_{lower}, \lambda - \lambda_{lower}) + A_{upper}\, L(\gamma_{upper}, \lambda - \lambda_{upper}) + A_{bare}\, L(\gamma_{cav}, \lambda - \lambda_{cav}) \quad (1)$$

where $L(\gamma,\Delta) = \gamma/\pi[\gamma^2 + \Delta^2]$ is a normalized Lorentzian function with linewidth $\gamma$ and detuning $\Delta$ from the center wavelength. We define the linewidths of the lower polariton, upper polariton, and bare cavity as $\gamma_{lower}$, $\gamma_{upper}$, and $\gamma_{cavity}$ respectively, while $A_{lower}$, $A_{upper}$, and $A_{bare}$ give the amount of emission from the three peaks.

Fig. 3a plots the fitted values of $A_{lower}$, $A_{upper}$ and $A_{bare}$ for ABE as a function of pump power when the QD and cavity are on resonance. The plot shows that all three peaks are linearly increasing with pump power for the range of powers used in this experiment. This data suggests that we are below saturation for all pump intensities. We note that it was not possible to increase the pump power to higher levels and investigate the saturation regime for ABE, because at higher pump powers we observed a significant blue shift of the cavity resonance. This shift was caused by heating of the $N_2$ on the surface of the sample that was used to shift the cavity near resonance with the QD. As the $N_2$ heated it evaporated from the surface and the cavity returned to its original resonance.

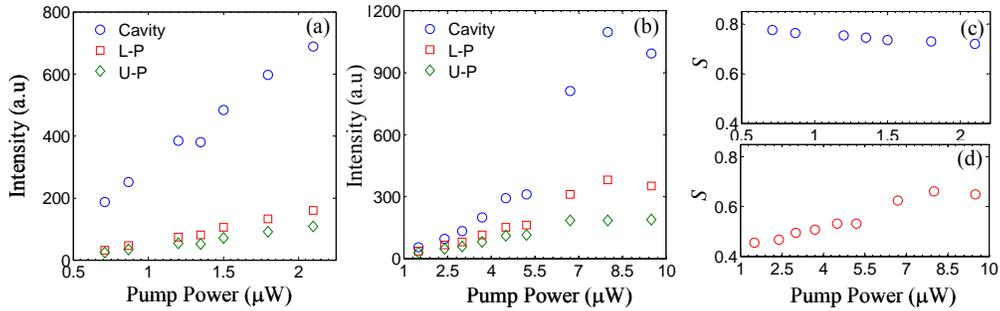

Fig. 3. (a and b) Integrated emissions from the cavity, lower (L-P) and upper (U-P) polariton states derived using a three-Lorentzian fit to the observed spectra at each value of the pump power for above-band (a) and quasiresonant (b) pumping. (c and d) Calculations of the derived emission ratio $S$ as a function of pump power for above-band (c) and quasiresonant (d) pumping.

In Fig. 3b we plot the areas under the three fitted Lorentzians as a function of pump power for QRE. At low pump powers the lower and upper polariton are linearly increasing, just as they did in ABE. At higher pump powers the emission from the two polaritons saturates. From the saturation data we determine the saturation power to be $P_{sat}$=8.5 µW. The reason that we could saturate the QD in QRE but not in ABE is because in QRE we are pumping below the bandgap of GaAs, so a significantly smaller fraction of the pump field is absorbed resulting in lower heating. We note that the two polariton states saturate to slightly different values. This discrepancy is attributed to the fact that the recorded spectra are slightly offset from the strong coupling point. Using detailed fitting, this detuning is inferred to be .02 nm which is close to the resolution of our spectrometer setup.

One interesting observation is that, unlike in ABE, the cavity emission in QRE is increasing nonlinearly even well below saturation. A likely explanation for this behavior is that in QRE the cavity feeding occurs primarily through the biexciton. Biexciton feeding plays a more prominent role in this pumping scheme because QD charging is not possible, so the only other state that the QD can fall into that is detuned from the cavity emission is the biexciton. Because biexciton emission requires the simultaneous absorption of two electron hole pairs instead of just one, creation of the biexciton state has a nonlinear pump power dependence [20, 22]. Near saturation, the polariton states begin to level off while the cavity emission continues to rise at an increasing rate, which is also consistent with cavity feeding being primarily driven by the biexciton.

The ratio of the bare cavity emission to the total emission is quantified by the ratio $S=A_{bare}/(A_{lower}+A_{upper}+A_{bare})$, which is plotted as a function of pump power for ABE in Fig. 3c. In ABE, the fraction of the emission due to the bare cavity is largely independent of pump intensity, achieving an average value of 0.75. This observation is consistent with recent theoretical predictions by Yamaguchi *et. al.* based on simulation of the QD-cavity master equation under conditions where electrons and holes are injected independently [17]. Under these conditions, the cavity feeding is mainly mediated by the charged exciton states, and in particular the positively charged exciton which is brighter and closer to cavity resonance. Both the charged exciton and single exciton emission are linearly proportional to pump power below saturation, as verified by experimental measurements on bare QDs [20, 22]. The cavity feeding should therefore also be linearly proportional to pump power, and the ratio of the bare cavity emission to polariton emission should consequently be pump-power independent. Thus, under ABE our measurements of bare cavity emission are well explained by cavity feeding through the charged exciton state.

Figure 3d shows the calculated parameter *S* as a function of power for QRE. Unlike ABE, in QRE *S* linearly increases with pump power below saturation. At 1.5 µW of pump power we measure *S* = 0.45, a 40% reduction relative to the average value for ABE below saturation. When the pump power is increased to 8 µW, *S* increases to 0.66. As the pump power approaches $P_{sat}$, *S* appears to saturate. The linear dependence of *S* below saturation is also consistent with theoretical predictions by Yamaguchi et. al. under the condition that electrons and holes are created in pairs inside the QD. In this case, QD charging is not possible and the primary mechanism for cavity feeding is through the biexciton. Below saturation, biexciton emission is expected to be a quadratic function of pump power, while single exciton emission should linearly increase with pump. Well below saturation the polariton emission should dominate the total spectrum so $S \approx A_{bare}/(A_{lower}+A_{upper})$ which is therefore the ratio of biexciton to exciton emission. This ratio should be a linear function of pump power. Thus, our observed measurement both for ABE and QRE are consistent with the prediction that bare cavity emission occurs due to other charge configurations of the QD, which feed the cavity due to the presence of pure dephasing.

We note that it has recently been shown by Hughes and Yao [27] that a spectral triplet feature could also occur due to the presence of in-plane losses. Such in-plane losses naturally exist due to the finite size of the photonic crystal, and can also be potentially induced by strong pump effects due to free-carrier absorption. In our measurements, we observe a change in the degree of feeding by simply changing the pumping scheme, which should not affect in-plane losses caused by the finite size of the photonic crystal. For pump induced losses such as free carrier absorption, we would expect to see a pump power dependence even well below QD saturation for ABE which generates many free carriers, but no pump power dependence for QRE because no free carriers are generated. Our data shows the opposite trend. Thus, although a spectral triplet can exist due to in-plane losses for certain photonic crystal devices, the spectral triplet we observe in our specific device appears to be of different origin and is better explained by cavity feeding from other QD charge configurations. It is likely that for our device the quality factor is heavily dominated by out-of-plane losses and therefore the effect of in-plane losses may not be observable. For devices with higher out-of-plane quality factors, this may not be the case.

## 6. Conclusion

In conclusion, we observe suppression in the bare cavity emission for a strongly coupled QD-photonic crystal cavity system through quasi-resonant excitation (QRE) as compared to above-band excitation (ABE). We also show that the ratio of bare cavity emission to polariton emission is mainly independent of pump power below saturation for ABE, but is linearly dependent for QRE. These measurements strongly support the theoretical predictions that the spectral triplet originates from off-resonant feeding of the cavity when the QD is in a different charge configuration, specifically the charged exciton and biexciton, through pure dephasing. QRE pumping is found to be a simple and effective way to reduce this effect, providing a more charge controlled environment for experiments in cavity quantum electrodynamics in the strong coupling regime.

**Acknowledgements**

This work was supported by a U.S. Army Research Office Young Investigator Award W911NF0710427, an NSF CAREER Award, the NSF Physics Frontier Center at the JQI, and the U.S. Army Research Office MURI award W911NF0910406.